\documentclass[10pt,twocolumn]{article}
\usepackage[margin=0.75in]{geometry}
\usepackage[numbers,sort&compress]{natbib}
\usepackage{url}
\usepackage{lineno}
\usepackage[T1]{fontenc}
\usepackage[utf8]{inputenc}
\usepackage{amsmath,amssymb}
\usepackage{booktabs}
\usepackage{graphicx}
\usepackage{siunitx}
\usepackage{subcaption}
\usepackage{xcolor}
\usepackage{tikz}
\usepackage{changepage}

\usetikzlibrary{positioning,fit,arrows.meta,backgrounds,calc,decorations.pathreplacing}

\title{Closing the realism gap in physics-based gait simulations \\ with a learned state prior}

\author{%
  Markus Gambietz$^{\ast}$ \and
  Zhihao Zhao \and
  Theodoros Balougias \and
  Xiang Wang \and
  Anne D.\ Koelewijn\\[4pt]
  \small Biomechanical Motion Analysis and Creation (BioMAC) group, \\
  \small Friedrich-Alexander-Universität Erlangen-Nürnberg (FAU), Erlangen, Germany \\
  \small *Corresponding author: \texttt{markus.gambietz@fau.de}
}
\date{}
\begin{document}

%TC:ignore
\makeatletter
\twocolumn[
  \begin{@twocolumnfalse}
    \maketitle
    \smallskip
    \bigskip

    \begin{abstract}
    \noindent 
    Predictive simulation of human movement is a promising tool for studying ``what-if'' scenarios in human movement and its underlying motor control, yet its realism is often limited. To address this gap, we incorporate a learned state prior that is trained on a large-scale dataset of human gait kinematics and external forces
    into predictive simulations. Resulting gait simulations yield kinematics and kinetics across diverse walking and running speeds that better match experimental data than current physics-based simulations, achieving accuracy comparable to data-driven models that reproduce learned data. Furthermore, our method enables robust hypothesis testing by demonstrating how varying optimality assumptions, muscle weakness, and footwear choices influence predicted gait. We also show that this prior generalizes well beyond its training data, successfully reconstructing full-body kinematics for curved running and cutting maneuvers from sparse marker sets. Ultimately, these results suggest that state priors should be broadly integrated into predictive simulations.
    \end{abstract}
    \bigskip
  \end{@twocolumnfalse}
]
\makeatother
%TC:endignore

%TC:ignore
%\begin{flushright}\footnotesize\input{wordcount.tex}\end{flushright}
%TC:endignore

\section{Introduction}

% P1: the big problem -- experiments say what happens, not why
Despite decades of research, we still cannot fully explain why humans move the way they do. Experiments capture movements, but answering \textit{what-if} questions, such as ``How would running economy change with different footwear?'', requires measuring every condition of interest, which is time-consuming and often expensive~\cite{hoogkamerComparisonEnergeticCost2018b}.
More \textit{fundamental} questions, for example whether humans choose their locomotion pattern based on energy efficiency or fatigue mitigation, cannot be observed directly in experiments
~\cite{ackermannOptimalityPrinciplesModelbased2010a,crowninshieldPhysiologicallyBasedCriterion1981}.

% P2: predictive simulation is the instrument that can answer them
Physics-based predictive simulation offers a different route: instead of measuring how a person moves, it computes how a person would move under a candidate hypothesis or scenario. Given a musculoskeletal model, a task, and an objective, it generates a motion that is physically consistent while optimising that objective, and it has been implemented using both reinforcement learning and optimal control~\cite{falisseRapidPredictiveSimulations2019d,schumacherEmergenceNaturalRobust2025a,parkGenerativeGaitNet2022,dhondtPredSimFrameworkRapid2024a}. Plausible proxies for the objective behind human movement exist: effort, which is related to fatigue mitigation, or metabolic energy efficiency~\cite{ackermannOptimalityPrinciplesModelbased2010a,crowninshieldPhysiologicallyBasedCriterion1981}. 
By changing the objective, the body model, or the environment, one can test hypotheses about human motor control and predict how humans would move under new conditions. %This makes predictive simulation, in principle, an instrument for testing hypotheses about human motor control.

% P3: the problem -- realism is entangled with the task
In practice, the motions that these simulations produce are not always realistic. Human motion has many degrees of freedom and many muscles to coordinate, so even a reasonable objective admits optima that no human would produce. Current methods mitigate this by adding hand-designed regularisation terms, such as penalties on joint accelerations or head movement~\cite{falisseRapidPredictiveSimulations2019d,veerkampEvaluatingCostFunction2021b,schumacherEmergenceNaturalRobust2025a}, or by tracking experimental data, i.e.\ by penalising deviation from measured kinematics or kinetics \cite{santosPredictiveSimulationPoststroke2021a,weissPairingPredictiveSimulations2024a}.
Both have drawbacks: hand-tuned terms do not yield realistic motion across the full range of speeds and gait modes, and tracking terms require experimental data for every new condition. Moreover, some conditions might not even be testable, which limits their use for prediction.
Data-driven motion generation models, on the other hand, produce highly realistic motion~\cite{tanGaitDynamicsGenerativeFoundation2026,magruderGaitEncoderFoundationModel2026,xie2024omnicontrol}, but their generated motion is conditioned on learned representations of recorded motion patterns. Therefore, analysing a change, e.g.\ in the ground contact stiffness, would still require experimental data for that condition.

\begin{figure*}[t!]
  \centering
  \includegraphics[width=0.87\linewidth]{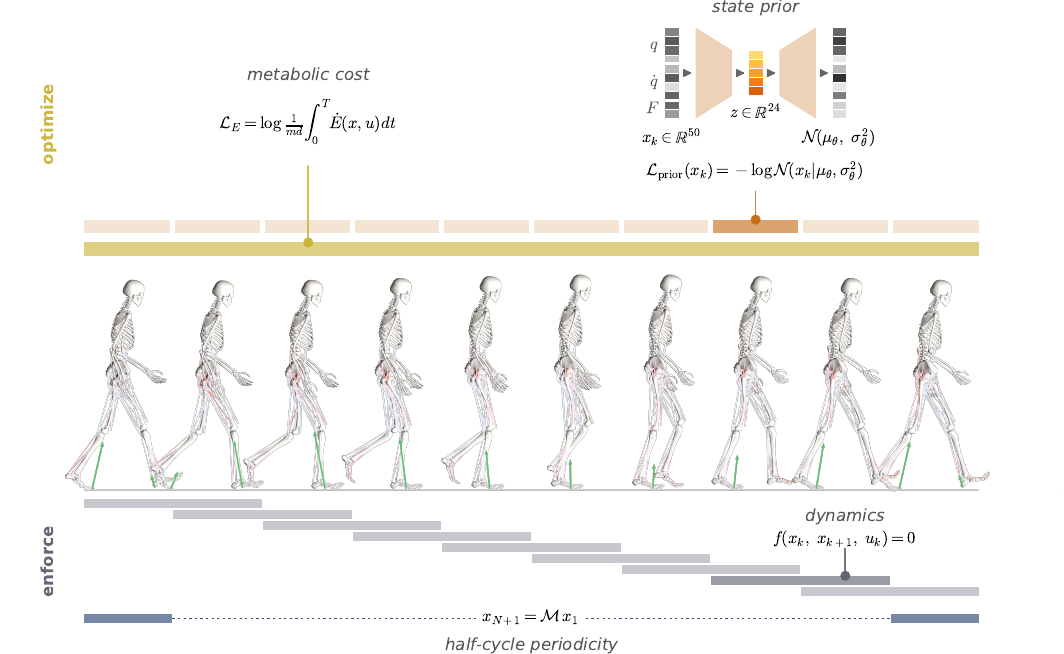}
  %\vspace{-1.5cm}
  \caption{\textbf{Overview of our simulation method.} Two objective terms are jointly optimised, a task objective (here: metabolic costs) and a learned state prior. Metabolic costs are estimated from the musculoskeletal model over the full gait cycle \cite{bhargavaPhenomenologicalModelEstimating2004a}, while the state prior is evaluated at every time step of the trajectory and then averaged. The resulting motion is constrained by the dynamics of the musculoskeletal model, which enforces that each state and control $x_k,u_k$ is physically consistent with the previous state~\cite{koelewijnBioMACSimToolboxMATLABToolbox2025b}.  Experiments 1 \& 2 use $N=50$ collocation nodes, which means that the temporal resolution is higher than in this schematic, while Experiment 3 uses \SIrange{149}{175}{} nodes.
  Experiments 1 and 2 also enforce half-cycle periodicity, which means that the state at node $N+1$ must be equal to the left-right mirrored first node state $\mathcal{M}x_1$.}
  \label{fig:vae-method}
\end{figure*}

\begin{figure*}[h!]
  \centering
  \includegraphics[width=\textwidth]{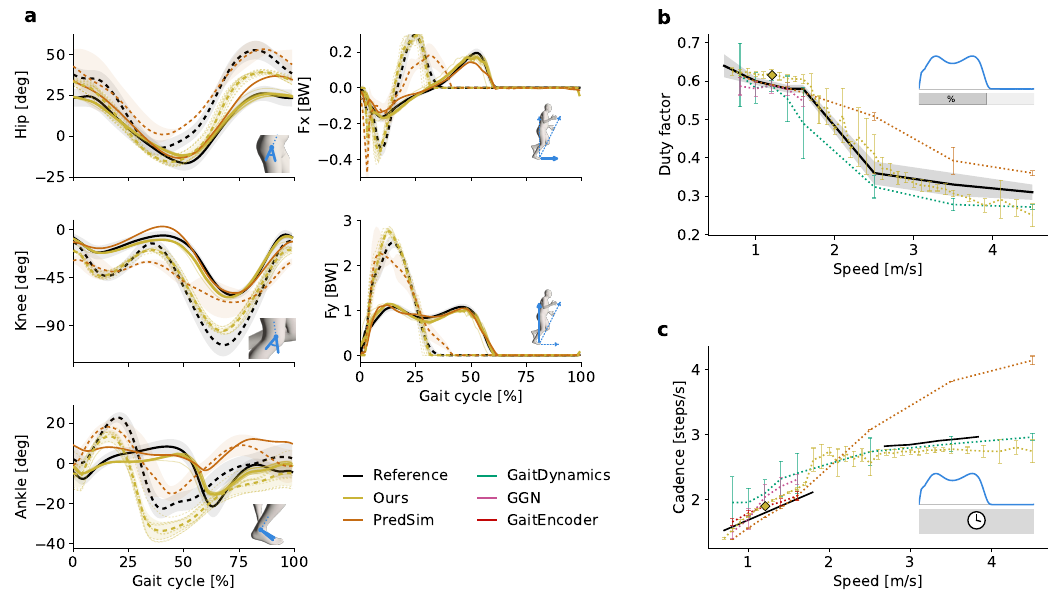}\\[6pt]
  \begin{minipage}[t]{0.5\textwidth}
    \centering
    \includegraphics[width=\linewidth]{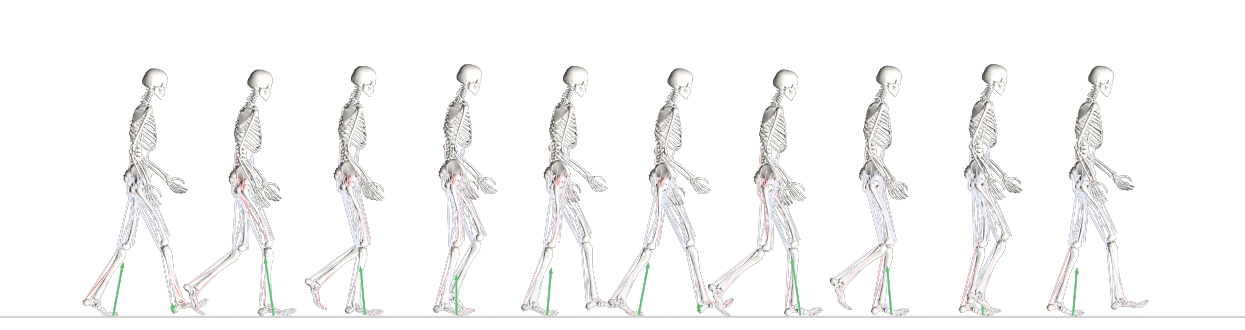}
  \end{minipage}%
  \begin{minipage}[t]{0.5\textwidth}
    \centering
    \includegraphics[width=\linewidth]{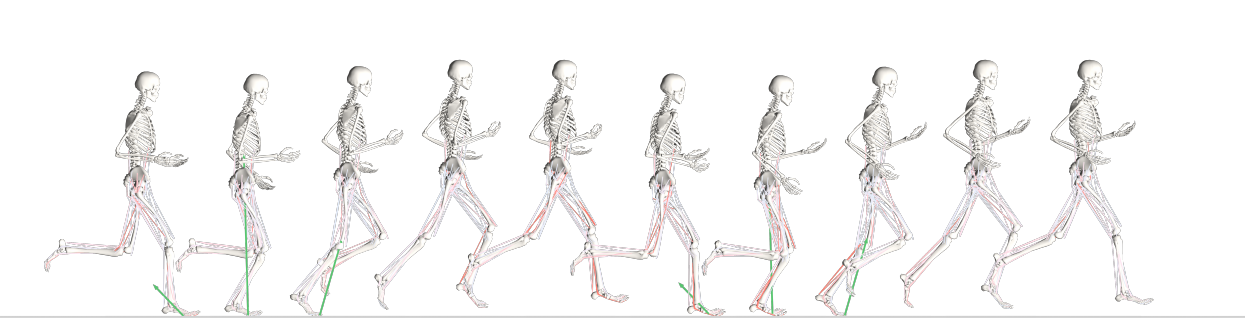}
  \end{minipage}\\[-2.5cm]
  \noindent\makebox[\textwidth][l]{\hspace*{0.4cm}{\sffamily\bfseries\fontsize{10pt}{12pt}\selectfont d}}\\[2.2cm]
  \caption{\textbf{State prior-based simulations match measured walking and running data.}
  \textbf{a)} Sagittal-plane hip, knee, and ankle joint angles (left) and fore-aft/vertical ground reaction forces (right) over one gait cycle, at a representative walking (\SI{1.2}{\meter\per\second}, solid) and running (\SI{3.5}{\meter\per\second}, dotted) speed. We find that our simulations are closer to experimental data~\cite{legoffPublicDatasetTreadmill2026,fukuchiPublicDatasetRunning2017} than current state-of-the-art simulations using PredSim \cite{dhondtPredSimFrameworkRapid2024a}. Individual trials are shown as thin lines, with the across-trial mean as a thick line and $\pm$ SD shaded. All curves are phase-aligned to the reference mean before averaging.
  \textbf{b)} Duty factor, which is the fraction of the gait cycle in stance, across every simulated speed. We compare ours to  reference data (mean $\pm$ SD shaded) \cite{fukuchiPublicDatasetRunning2017,legoffPublicDatasetTreadmill2026}, PredSim, GaitDynamics \citep{tanGaitDynamicsGenerativeFoundation2026}, and Generative GaitNet (GGN) \citep{parkGenerativeGaitNet2022}. The diamond marks our result for ``self-chosen'' walking speed. Our method yields a duty factor most consistently aligns with both walking and running reference data.
  \textbf{c)} Step cadence vs.\ speed for the same sources plus GaitEncoder \citep{magruderGaitEncoderFoundationModel2026}, overlaid with literature reference relations (walking: \citep{hansenRoleStrideFrequency2017}, running: \citep{leacoxEffectRunningSpeed2025}). Here, current methods cannot reproduce the reference cadence-speed relation across the whole speed range. The diamond again marks our self-chosen speed result.
  \textbf{d)} Renders of exemplary gait cycles at walking (\SI{1.2}{\meter\per\second}) and running (\SI{4.3}{\meter\per\second}) speeds.}
  \label{fig:transition}
\end{figure*}

\begin{figure}[t]
  \centering
  \includegraphics[width=\linewidth]{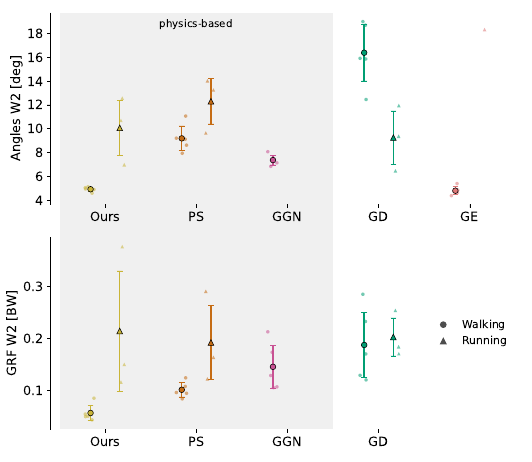}
  \caption{\textbf{State prior-based simulations outperform current physics-based kinematics and kinetics, matching non-physical baselines.} Distribution error (W2 Wasserstein distance) of simulated compared to reference kinematics for walking and running for our method, PredSim (PS), Generative GaitNet (GGN), GaitDynamics (GD), and GaitEncoder (GE). Each dot represents the Wasserstein distance between the simulation and reference distribution for one speed, with the mean $\pm$ SD across speeds shown as a point or triangle with error bars. Lower is better.}
  \label{fig:baselines}
\end{figure}

\begin{figure*}[t!]
  \centering
  \includegraphics[width=\textwidth]{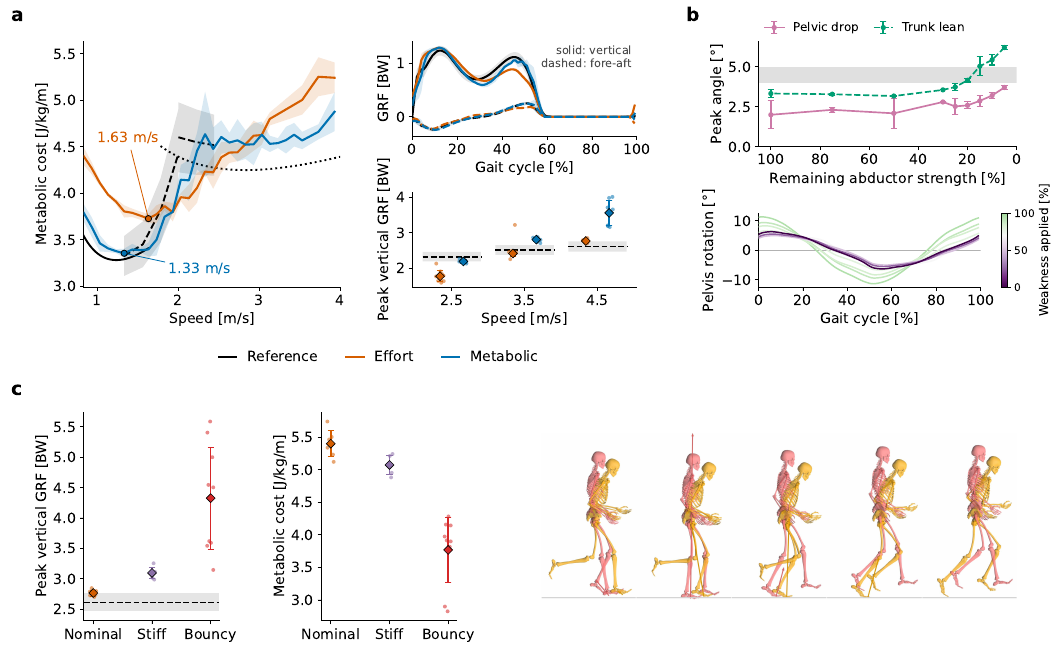}
  \caption{\textbf{State prior-based simulations enable hypothesis testing in gait.} 
  %\textbf{a)} and \textbf{b)} are evaluating optimality criteria in human gait, where both effort and metabolic cost are considered as plausible objective functions. 
  \textbf{a)}~compares metabolic cost and effort as the task objective, both are considered as plausible objective functions. 
  When using effort instead of metabolic costs, we show that the estimated metabolic costs using Umberger's model~\cite{umbergerStanceSwingPhase2010b} don't match the reference patterns (solid: \cite{ralstonEnergyspeedRelationOptimal1958}, dashed: \cite{abeWhyWeTransition2019}, dotted: \cite{batlinerDoesMetabolicRate2018}) with the minimum metabolic cost appearing at a higher speed (left). We also find that the push-off force during walking is too low (top right, data at 1.6 m/s), and that the peak vertical ground reaction force is too small for low running speeds, but more realistic for higher running speeds (bottom right). One outlier trial at 4.5 m/s was removed (Metabolic, $\approx$\SI{5.5}{} BW).
  %\textbf{b)} compares walking at unconstrained speeds using different metabolic cost models (Bhargava (BHAR)~\cite{bhargavaPhenomenologicalModelEstimating2004a}, Umberger (UMBE)~\cite{umbergerStanceSwingPhase2010b}, Lichtwark (LICH)~\cite{lichtwarkModifiedHillMuscle2005a}, Houdijk (HOUD)~\cite{houdijkEvaluationHillBased2006a}) and shows that emergent walking speeds differ by up to \SI{0.15}{\meter\per\second}, with Umberger and Bhargava producing the most realistic speed.
  \textbf{b)} simulates walking with weak hip abductors. Top: we show the peak pelvic drop and trunk lean angles increase with decreasing abductor strength, both of which are commonly observed in patients with weak hip abductors~\cite{krautwurstInfluenceHipAbductor2013}. Both are considered abnormal above \SIrange{4}{5}{\degree} (annotated by the shaded area). Bottom: we show that the pelvis rotation increases with decreasing abductor strength. 
  \textbf{c)} investigates the effect of ground contact stiffness and damping on running gait at \SI{4.5}{\meter\per\second}. ``Stiff'' refers to a four-times-stiffer ground contact than the default, while ``Bouncy'' is a ground contact model without damping. Both variants increase the peak vertical ground reaction force and reduce metabolic costs. Two solutions from the ``Bouncy'' condition led to plyometric running, which is not observed in humans and characterised by very short ground contact times and exceptionally low metabolic costs.}
  \label{fig:designtool}
\end{figure*}

% P4: the alternative -- learned models of human movement, and how ours differs

Integrating learned models of human movement into physics-based simulations provides an alternative to hand-designed terms and to reference tracking. Such models have been used to stabilise pose estimation~\cite{rempeHuMoR3DHuman2021a,SMPL-X:2019} and, notably, as a training reward in reinforcement learning-based physical simulation~\cite{pengAMPAdversarialMotion2021,fengMuscleVAEModelBasedControllers2023,parkMuscleSynergyPriors2026,weiScalingWholeBodyHuman2026,simosKINESISMotionImitation2026}. 
We apply the same idea to optimal control-based predictive simulation, where a learned probability function replaces hand-designed regularisation terms. Compared to a trained policy, a trajectory optimisation returns a local optimum of the problem we pose. Differences between simulations can therefore be attributed to the change in objective, body model, or environment, rather than testing the generalization of the training process.

% P5: what the prior is
We refer to the probability estimate as a \emph{state prior}: a smooth, differentiable reconstruction probability over the states we can measure, covering joint angles and velocities as well as ground reaction forces. Essentially, the state prior encodes how likely a given state is to be observed in human gait.
We obtain it by training a probabilistic autoencoder once on a large motion-capture dataset~\cite{werlingAddBiomechanicsDatasetCapturing2025} and then keeping it fixed across all simulations. We evaluate the state prior at every time step of a trajectory optimisation problem solved with numerical methods like direct collocation and use it as an ordinary objective term, alongside, for example, a term for metabolic energy (see Fig.~\ref{fig:vae-method}). Compared to a tracking term that penalises deviation from a specific reference motion, the state prior is agnostic to the specific motion and does not prescribe a particular trajectory, speed, or gait mode. The state prior also only sees individual states, not the full trajectory, so it does not impose a specific gait speed or cadence. 
Within the optimisation, the physics then determines which motions are possible, the task objective determines which of those motions is preferred, and the state prior guides the solution towards the states that humans likely adopt.

% P6: what we demonstrate
% Keeping the prior fixed while varying the hypothesised objective, the body model, or the physical environment allows us to examine which aspects of human gait each hypothesis reproduces. 
We demonstrate this approach in three experiments. In the first, predictive simulations across the full range of walking and running speeds reproduce measured kinematics and ground reaction forces ($r=0.94$--$0.98$) more closely than existing physics-based simulations, and comparably to data-driven models that are trained to reproduce the data. Realistic cadences, a self-chosen walking speed, and the walk-to-run transition emerge from these simulations without being imposed by the objective function or by reference data. In the second experiment, we test various hypotheses, showing that the prior does not overconstrain the solution space, but instead allows us to examine how different optimality criteria, muscle weakness, and ground contact models affect the predicted gait, including solutions that are physically valid but are not observed in humans. In the third, we show that the prior generalises beyond its training data by reconstructing full-body kinematics of curved running and cutting manoeuvres from sparse marker sets. In conclusion, the same framework applies wherever the question is how a person would move under conditions that have not been measured, for example in footwear or assistive-device design.

\section{Results}

\begin{figure*}[t]
  \centering
  \includegraphics[width=\textwidth]{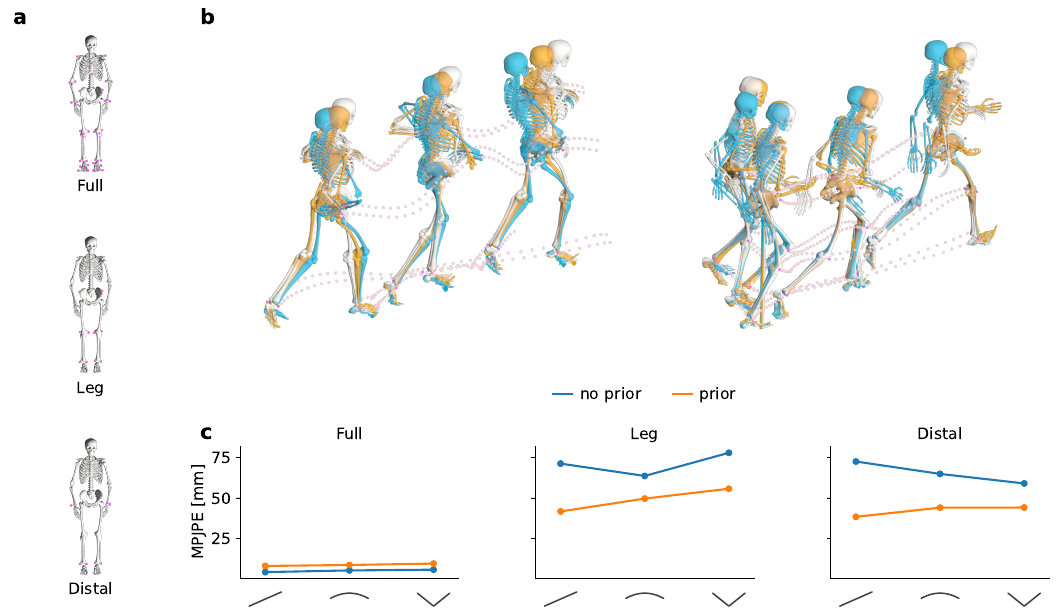}
  \caption{\textbf{State prior-based simulations enable sparse tracking and reconstruction.}
  \textbf{a)} Marker sets used for tracking: \emph{Full} (42 markers), \emph{Leg} (11 markers on the pelvis, knees, and ankles), and \emph{Distal} (8 markers on the ankles and hands).
  \textbf{b)} Exemplary motions reconstructed with and without the state prior: straight running tracked with the \emph{Distal} marker set (left) and a cutting manoeuvre tracked with the \emph{Leg} marker set (right). Grey shows the inverse kinematics reference, orange the reconstruction with the prior, and blue the reconstruction without it.
  \textbf{c)}~Mean per-joint position error (MPJPE) with respect to the inverse kinematics reconstruction, for each marker set and motion (straight running, curved running, cutting manoeuvre), with and without the prior.
  }
  \label{fig:sparse}
\end{figure*}

\subsection*{State prior simulations yield realistic human gait across the full speed range and a self-chosen walking speed}

We run predictive simulations across the full range of walking and running speeds, from \SIrange{0.73}{5.53}{\meter\per\second}. In these simulations, we optimise a combination of a task objective (metabolic cost) and the state prior, while enforcing physical consistency, half-cycle periodicity and a target speed. Across all speeds with reference data (\SIrange{0.8}{1.6}{\meter\per\second} for walking \cite{legoffPublicDatasetTreadmill2026}, \SIrange{2.5}{4.5}{\meter\per\second} for running \cite{fukuchiPublicDatasetRunning2017}), our simulations achieve correlations of $r=0.94$--$0.98$ for kinematics and ground reaction forces.
Not only are the gait patterns realistic, but a natural cadence and the walk-to-run transition also emerge naturally from the simulation. In Figure~\ref{fig:transition}b, the walk-to-run transition is characterised by the duty factor, which is the fraction of the gait cycle in stance. In humans, it is observed that the duty factor is around \SI{0.6}{} for walking and less than \SI{0.4}{} for running, with a transition between the two at around \SI{2}{\meter\per\second}~\cite{winterBiomechanicsMotorControl2009b,abeWhyWeTransition2019}.
Current methods, on the other hand, fail to reproduce the reference cadence-speed relation across the whole speed range, whereas our method does (Fig.~\ref{fig:transition}c).
We compared the distributions of our solutions to the reference data using the Wasserstein distance (Fig.~\ref{fig:baselines}). For walking kinematics, our simulations reach \SI{4.96(43)}{\degree}, against \SI{9.19(104)}{\degree} for optimal control-based PredSim \cite{dhondtPredSimFrameworkRapid2024a}, \SI{7.37(41)}{\degree} for reinforcement learning-based Generative GaitNet \cite{parkGenerativeGaitNet2022} and \SI{16.39(237)}{\degree} for data-driven GaitDynamics \cite{tanGaitDynamicsGenerativeFoundation2026}, and are on par with GaitEncoder \SI{4.80(34)}{\degree} \cite{magruderGaitEncoderFoundationModel2026}, which is a data-driven method trained to reproduce the data. For running kinematics, we reach \SI{10.08(238)}{\degree}, compared to \SI{12.31(191)}{\degree} for PredSim, and \SI{9.26(224)}{\degree} for GaitDynamics, while GaitEncoder is not meant to produce running gaits.
The same holds for the ground reaction forces, where we reach \SI{0.084(26)}{BW} for walking and \SI{0.199(97)}{BW} for running, against \SI{0.102(14)}{} and \SI{0.193(71)}{BW} for PredSim and \SI{0.188(63)}{} and \SI{0.203(36)}{BW} for GaitDynamics, and \SI{0.146(42)}{BW} for Generative GaitNet during walking. Thus, none of the  baselines are reproducing spatiotemporal, kinematic, and kinetic features for both walking and running as closely as our method. 
% Note: n per cell -- walking 5 speeds, running 3 speeds; GGN produces no running solutions, GE no ground reaction forces, and the GE running value is a single speed (n=1).

The simulations were repeated with one of the two objective terms removed at \SI{1.2}{} and \SI{3.5}{\meter\per\second}. Removing the prior increases the kinematic distribution error by a factor of \num{2.6} to \num{4.5}, whether the task objective is effort (walking: \SI{13.01}{\degree}; running: \SI{28.61}{\degree}) or metabolic cost (\SI{22.13}{\degree}; \SI{35.29}{\degree}), against \SI{4.94}{} and \SI{10.50}{\degree} for the full objective. Removing the task objective instead leaves the kinematic error comparable to the full objective (\SI{6.87}{\degree}; \SI{9.46}{\degree}), while the error in the ground reaction forces is two to three times higher, at \SI{0.168}{} and \SI{0.209}{BW} against \SI{0.051}{} and \SI{0.112}{BW} in walking and running, respectively. Neither ablation reproduces both the kinematics and the ground reaction forces as closely as the full objective. 

We also ran simulations in which we left the speed unconstrained, so that the walking speed is determined by the objective. We minimise metabolic cost, and the walking speed that emerges, \SI{1.21(2)}{\meter\per\second}, is close to the energetically optimal walking speed of \SI{1.23}{\meter\per\second} measured by~\citet{ralstonEnergyspeedRelationOptimal1958}. Humans often walk faster than their energetic optimum and self-select speeds between \SIrange{1.2}{1.5}{\meter\per\second}~\cite{brinkerhoffWordsMatterInstructions2022}, so our solutions lie at the lower end of that range.
The baselines cover a much wider range. PredSim walks at \SI{0.75(9)}{\meter\per\second} and Generative GaitNet at \SI{1.09(18)}{\meter\per\second}, both below the range that humans self-select. In unconstrained settings, GaitDynamics does not favour a particular speed, but reproduces a range, spanning \SIrange{0.72}{3.37}{\meter\per\second} and thus covering speeds representable for their data set. GaitEncoder, when set to produce healthy gait, emerges at \SI{1.24(15)}{\meter\per\second}. Ours and GaitEncoder therefore both fall inside the self-selected range, as does the model-based reinforcement learning approach of \citet{suSimulatingHumanWalking2023}, which reports a self-selected walking speed of \SI{1.45}{\meter\per\second}.

% --- Ablations (1.2 and 3.5 m/s) ---
% W2, angles [deg] / GRF [BW], at 1.2 and 3.5 m/s:
%   Ours (metabolic + prior)   4.94 / 0.051   10.50 / 0.112
%   effort + prior             4.51 / 0.086   12.47 / 0.088
%   prior only (_none)         6.868 / 0.168   9.464 / 0.209   (final; n=8 and n=9)
%   effort, no prior           13.01 / 0.168  28.61 / 0.610
%   metabolic, no prior        22.13 / 0.092  35.29 / 0.169
% --- end ablations ---

\subsection*{State prior-based simulations enable hypothesis testing in gait}

% Source data (2026-08-31, MG):
%   Walk-to-run transition: metabolic 1.74 m/s, effort 2.16 m/s
%   Free-speed by metabolic model (mean +- SD, m/s):
%     BhargavaAct 1.2075 +- 0.0205 | Houdijk 1.1452 +- 0.0190 | Umberger 1.2860 +- 0.0179
%     Lichtwark   1.1461 +- 0.0458 | 30bwa_bhargavaact 1.1979 +- 0.0082 | 30bws_bhargavaact 1.2213 +- 0.0304
%   Ground contact: nominal / 4x stiffer / no damping; effort used as task objective

% --- Experiment 2a: effort vs metabolic cost ---
The choice of objective and musculoskeletal model changes the outcome of gait simulations (Fig.~\ref{fig:designtool}). First, we tested the effect of switching the metabolic cost to an effort-based task objective, which aims to minimize cubed muscle activations. When optimizing for metabolic costs, we find that the resulting energy landscape over walking and running speeds qualitatively matches experimental data~\cite{ralstonEnergyspeedRelationOptimal1958,abeWhyWeTransition2019,batlinerDoesMetabolicRate2018}, while effort-optimization predicts a higher optimal walking speed at \SI{1.63}{\meter\per\second} and does not reproduce the characteristic double U-shape of the metabolic cost curve (Fig.~\ref{fig:designtool}a, left). 
The two objectives also differ in the ground reaction forces they produce. During walking, effort underestimates the push-off peak by \SIrange{11.8}{20.6}{\percent} across speeds (on average \SI{0.919}{BW} against a measured \SI{1.079}{BW}), whereas metabolic cost stays within \SIrange{5.8}{6.3}{\percent} (\SI{1.014}{BW}).
During running, effort yields lower vertical ground reaction force peaks than metabolic cost at all three simulated speeds. At \SI{2.53}{\meter\per\second} it underestimates the measured peak of \SI{2.324(133)}{BW} by \SI{26}{\percent} (\SI{1.717(133)}{BW}), while metabolic cost stays within \SI{6}{\percent} (\SI{2.186(41)}{BW}).
At the two higher speeds this reverses: at \SI{3.53}{\meter\per\second} effort gives \SI{2.476(247)}{BW} against a measured \SI{2.512(138)}{BW}, and at \SI{4.53}{\meter\per\second} \SI{2.746(57)}{BW} against \SI{2.611(147)}{BW}, whereas metabolic cost overestimates the peak by \SI{10}{\percent} and \SI{33}{\percent} (\SI{2.771(73)}{BW} and \SI{3.481(599)}{BW}; Fig.~\ref{fig:designtool}a, left).
Across all walking gaits, metabolic cost describes walking more closely than effort, both in kinematics (MAE \SI{2.74}{\degree} against \SI{3.83}{\degree}) and in ground reaction forces (\SI{0.030}{BW} against \SI{0.052}{BW}), while for running the kinematic errors are comparable (\SI{9.05}{\degree} against \SI{9.14}{\degree}) and effort reproduces the ground reaction forces more closely (\SI{0.081}{BW} against \SI{0.106}{BW}).
Neither objective therefore reproduces the whole speed range: metabolic cost describes walking and running at low speeds better, while effort better describes running at higher speeds.

% --- Experiment 2b: metabolic cost models ---
Within the metabolic cost hypothesis, the choice of energy model shifts the emergent walking speed. Umberger's model \cite{umbergerStanceSwingPhase2010b} produces the fastest walking at \SI{1.29(2)}{\meter\per\second}, Bhargava's model \cite{bhargavaPhenomenologicalModelEstimating2004a} at \SI{1.21(2)}{\meter\per\second}, and the models of Houdijk \cite{houdijkEvaluationHillBased2006a} and Lichtwark \cite{lichtwarkModifiedHillMuscle2005a} the slowest at \SI{1.14(2)}{} and \SI{1.15(5)}{\meter\per\second}. The emergent speeds therefore differ by \SI{0.14}{\meter\per\second} between models. 
% TODO (MG to decide later): whether to keep the comparison to the self-selected gap in the sentence above.

We also investigated effects of changing the musculoskeletal model, as seen in Fig.~\ref{fig:designtool}b and \ref{fig:designtool}c. First, we simulated walking with weak hip abductors. Weak hip actuators are commonly observed to cause Trendelenburg gait, where excessive pelvic drop, or trunk lean as a compensation, is observed during stance~\cite{krautwurstInfluenceHipAbductor2013,kendallStepsValidationTrendelenburg2013}. In our simulations, we find that hip abductors saturate during the stance phase when their strength is reduced to less than \SI{30}{\percent} of the nominal value. Up to \SI{30}{\percent} remaining strength, only small changes in pelvic drop and trunk lean are observed, while walking speed decreases from \SI{1.21}{\meter\per\second} to \SI{1.16}{\meter\per\second} and step width increases from \SIrange{11}{14.3}{\centi\meter}. When the remaining strength is reduced further, compensation strategies emerge, with trunk lean reaching up to \SI{6.2}{\degree}. Pelvic drop increases to a maximum of \SI{3.7}{\degree}. The compensation mechanisms creates a waddling gait, with pelvis rotations exceeding \SI{10}{\degree}. Here, the metabolic cost optimization predicts that weaker hip actuators lead to faster walking speeds with lower step widths up to \SI{1.38}{\meter\per\second} and \SI{8.4}{\centi\meter}, respectively. Metabolic cost, however, rises steadily with decreasing abductor strength, from \SI{3.35}{\joule\per\kilo\gram\per\meter} at \SI{100}{\percent} strength to \SI{3.78}{\joule\per\kilo\gram\per\meter} at \SI{25}{\percent} strength, and then more steeply to \SI{4.51}{\joule\per\kilo\gram\per\meter} at \SI{5}{\percent} strength. 

% --- Experiment 2c: ground contact model ---
To simulate the effects of different running shoes, we ran running simulations at \SI{4.5}{\meter\per\second} with the nominal Hunt-Crossley ground contact model, a four-times-stiffer model, and a model without damping. We used effort as the task objective as it better describes fast running.
The stiffer contact model increases the peak vertical ground reaction force from \SI{2.746(57)}{} to \SI{3.079(100)}{BW} and decreases the metabolic cost from \SI{5.40}{} to \SI{5.08}{\joule\per\kilo\gram\per\meter}.
Removing damping has a much larger effect, raising the peak force to \SI{4.250(850)}{BW} and lowering the metabolic cost to \SI{4.03}{\joule\per\kilo\gram\per\meter}.
Two of these solutions converge to plyometric running, characterised by very short ground contact times, high impact forces, and a small kinematic range of motion. Their knee range of motion is \SI{50.3}{} and \SI{60.5}{\degree}, against \SI{79.1(56)}{\degree} for the remaining solutions without damping, and their metabolic cost is only \SI{2.86}{\joule\per\kilo\gram\per\meter}.
These solutions are better in terms of effort, but the state prior scores them as less likely than the nominal ones (\SI{-2.03}{} against \SI{-2.44}{}), and plyometric running is not observed in humans. 

\subsection*{The state prior enables reconstruction from sparse marker sets}

We reconstructed straight running, curved running, and a cutting manoeuvre from three marker sets: the full set of 42 markers, a leg set of 11 markers on the ankles, knees, and pelvis, and a distal set of 8 markers on the ankles and hands (Fig.~\ref{fig:sparse}a).
With the full marker set, adding the prior increases the mean per-joint position error (MPJPE) from \SI{5.3}{} to \SI{8.9}{\milli\meter}. %The markers already determine the motion here, so the prior only pulls a well-constrained solution towards the states it considers likely.
With the reduced marker sets, the prior instead lowers the error: from \SI{70.8}{} to \SI{49.0}{\milli\meter} for the leg set and from \SI{65.4}{} to \SI{42.2}{\milli\meter} for the distal set (Fig.~\ref{fig:sparse}c). The reduction is consistent across all six combinations of marker set and motion, ranging from \SI{13.9}{\milli\meter} for curved running with the leg set to \SI{34.0}{\milli\meter} for straight running with the distal set.
%With the prior, the eight distal markers reconstruct the motion more accurately (\SI{42.20}{\milli\meter}) than the eleven leg markers do without it (\SI{70.83}{\milli\meter}).
Curved running and the cutting manoeuvre are not represented in the training data of the prior, so this also shows that the prior transfers to motions it has not seen.

%\vspace{5.3em}

\section{Discussion}

% D1: summary -- what the three experiments show together
In this work, we proposed integrating a learned state prior into optimal control simulations of human gait. Our evaluation demonstrates that this approach successfully reproduces measured kinematics and ground reaction forces across a broad spectrum of walking and running speeds, outperforming previous physics-based simulation methods. By achieving a high degree of realism in both kinematics and kinetics, this framework provides a reliable foundation for predictive modeling under altered conditions. 
Based on the greater realism of the simulations, we can more confidently test hypotheses about human movement by varying the objective, body model, or environmental constraints. Using state prior-based simulations, we found that a metabolic cost objective better describes walking and running at low speeds, while an effort objective better describes running at higher speeds. We also tested alterations to the musculoskeletal model, showing how changing the ground contact stiffness and damping affects the resulting gait. Finally, we demonstrated that the state prior can be used to reconstruct full-body kinematics from sparse marker sets, even for motions absent from the training data. 

Realistic simulations are a precondition for trust and downstream applications. If our simulations would not reproduce measured kinematics and kinetics, they cannot be reliably trusted to generalize to unmeasured conditions or to predict the effects of interventions. The results of our first experiment show that our method reproduces measured kinematics and kinetics across the full range of walking and running speeds more closely than previous physics-based simulations, and comparably to data-driven models that reproduce their training data without guarantees for physical plausibility. This allows us to use the same framework for hypothesis testing and motion reconstruction in the subsequent experiments.

In the second experiment, we first tested whether a metabolic cost or a cubic effort objective better describes human gait, i.e. the question whether humans plan their movement to minimize energy expenditure or to minimize fatigue. Which hypothesis holds for human walking is an open question, and previous work for walking and lower running speeds has proposed that effort \cite{ackermannOptimalityPrinciplesModelbased2010a}, metabolic cost \cite{andersonDynamicOptimizationHuman2001a}, or a combination of both \cite{afschriftBenchmarkingPredictiveCapability2025,falisseRapidPredictiveSimulations2019d} may be correct. We found that neither objective reproduces the whole speed range, as metabolic cost describes walking and running at low speeds better, while effort better describes running at higher speeds. Compared to previous predictive simulations, where solutions sometimes differed vastly between objectives~\cite{ackermannOptimalityPrinciplesModelbased2010a}, our simulations show a more localized effect of the objective, with both formulations producing human-like walking patterns closer to measured data than previous work. % (W2 angles: \SI{5.15}{\degree}, GRF: \SI{0.099}{\degree}). 
Thus, our simulations do not support the fatigue-minimization hypothesis for walking for two reasons: First, compared to the metabolic cost hypothesis, the effort objective yields a \SI{53}{\percent} lower peak in the triceps surae activation, which fails to reproduce the measured push-off peak forces but requires \SI{9.2}{\percent} more positive mechanical power over all joints. Secondly, the effort objective predicts a higher optimal walking speed than measured, which, however, is consistent with previous state-of-the-art predictive simulations using a mixed effort-metabolic objective~\cite{falisseRapidPredictiveSimulations2019d}. Similarly, we found that effort minimization leads to a duty factor overestimation in slow running as observed in previous simulations \cite{dhondtPredSimFrameworkRapid2024a}. 

A related question is whether the choice of metabolic cost model matters, and while the emergent walking speed differs by \SI{0.14}{\meter\per\second} between models, the kinematic and kinetic patterns are similar between them. Therefore, which metabolic cost model is used is likely less important, and we chose to use Bhargava's model for the practical reason that it converged more reliably and faster than the other models, and to use Umberger's model to evaluate metabolic costs post-simulation because same-model evaluations usually result in underestimations~\cite{afschriftBenchmarkingPredictiveCapability2025}. 
On that note, effort would be a more convenient objective to use, as it requires on average \SI{2.3}{\times} fewer iterations to converge than metabolic cost, with solve times being \SI{43.2}{\minute} per simulation against \SI{136.7}{\minute} for metabolic cost, but the results show that it should not be used for predictive simulations of walking and running at low speeds. Our reported runtimes are longer than those reported by PredSim~\cite{dhondtPredSimFrameworkRapid2024a}, which is due to the initial guess being a standing pose, while PredSim uses a walking reference motion. When using a walking reference motion at \SI{1.33}{\meter\per\second} as initial guess, simulations take on average \SI{47.7}{\minute} for metabolic cost.

Next, we tested the effect of changing the musculoskeletal model rather than the objective, by varying the muscle strengths, as well as ground contact stiffness and damping. When testing the effect of weak hip abductors, we found that the simulations reproduce the expected compensated Trendelenburg gait, with increased trunk lean and pelvis rotation range of motion, but less exaggerated pelvic drop. As the simulation finds optimal patterns for the given musculoskeletal model, it already predicts that humans would compensate for weak hip abductors. 
The onset of the compensation mechanisms occurs at \SI{30}{\percent} remaining abductor strength, and passes \SI{5}{\degree} trunk lean at \SI{15}{\percent} remaining strength. With \SI{20}{\percent} remaining strength, the results only point weakly towards a compensated gait, which is consistent with \citet{kendallStepsValidationTrendelenburg2013}'s observation that this amount of remaining strength is sufficient to pass the Trendelenburg test. In their study, they found only one participant to show a compensated gait, whose hip abductors were similar in strength to our \SI{10}{\percent} remaining strength condition. However, the simulations predict that the compensation strategy involves a faster walking speed and narrower step width, which is not observed in humans. This is likely because the simulations are optimising for metabolic cost, but humans with weak hip abductors need to consider other factors than merely metabolic costs. With the simulated compensation strategy, the hip abductors are still saturated during stance, which could lead to pain or discomfort, and humans would also need to consider stability and fear of falling, which is not captured by the metabolic cost objective. This limitation shows that it is important to be able to solve free-speed simulations, as constraining the speed would have hidden this effect, and leads to the open question in which way human motion planning differs between healthy and impaired individuals \cite{falisseRapidPredictiveSimulations2019d}.

We found that stiffer and less damped ground contact raised the impact peak and lowered metabolic cost, and that without damping, non-human-like plyometric running emerged. We deliberately introduced large changes to the ground contact model to test whether the state prior-driven simulations would still produce human-like gait and how the resulting gait would change. While the stiffer model lead to no substantial changes in gait kinematics, it resulted in a higher peak vertical ground reaction force and lower metabolic cost, which is opposite to what is observed in experiments~\cite{ferrisMidsoleFoamCompliance2026,baltichIncreasedVerticalImpact2015}.
Here, the nominal ground contact stiffness is set to \SI{85}{N/mm}, which is in the range of measured stiffnesses of running shoes of \SIrange{67}{117}{N/mm}~\cite{fohrmannBiomechanicalFactorsAssociated2026}, but the 4x-stiffer condition is far outside of that range. Therefore, while humans usually adjust their leg stiffness to compensate for changes in surface stiffness so that the effective stiffness remains roughly constant~\cite{shortenEnergyReturnFootwear2024}, the simulations only partially reproduce this behaviour as we overexaggerated the change in ground stiffness, and the vertical stiffness is increased by \SI{10.8}{\percent} in the 4x-stiffer condition. It has to be noted that the foot-shoe interface is not modelled, so additional losses will occur, and the foot itself will not get to 100\% energy return, thus the \SI{5.9}{\percent} reduction in metabolic cost is likely an overestimation. Conversely, stiffer midsoles actually increase metabolic costs in experiments~\cite{ferrisMidsoleFoamCompliance2026}. These results are limited by the specific ground contact model, as it is known to vastly overestimate the energy loss during ground contact~\cite{bainesExperimentalEstimationEnergy2018}. 

When idealizing the energy return of the ground contact model by removing damping, some simulations converged to a non-human-like plyometric running gait, which is characterised by short ground contact times, high impact forces, and a small knee range of motion. Plyometric running also appeared when using the ground contact model of \citet{nitschkeEfficientTrajectoryOptimization2020d}, especially when using the default musculoskeletal model instead of the inertial parameters from \citet{gambietzBodyHullsMusculoskeletal2026}.
While the prior scores these solutions as less likely than the nominal ones, it does not exclude them, thus showing that the prior does not overconstrain the solution space, and that if we ask a question about unrealistic conditions, unrealistic gait emerges. As the reduction in metabolic cost is substantial, we tested further whether this running style could actually provide any competitive advantage and therefore simulated running at \SI{6.0}{\meter\per\second}, a world-class marathon pace. The resulting peak vertical ground reaction force is \SI{5.82(85)}{BW} and the overall peak activations are ranging from \SIrange{190}{225}{\percent}. Both values might be withstandable for short durations, as the impact forces are similar to drop landing from a height of approximately \SI{0.75}{\meter} height~\cite{niuPeakVerticalGround2014}, and it is already common practice to double the force capacity in musculoskeletal models when simulating running~\cite{haralabidisSimulationsRevealHow2025,arnoldHowMuscleFiber2013}. Thus, the hypothesis that plyometric running could be a competitive strategy is not ruled out by the simulations, but it is likely not a sustainable strategy for long-distance running. However, it should be tested in experiments. If the strategy proves infeasible in practice, it would reveal a failure mode that the prior does not capture, for example pain, discomfort or injury risk. These failure modes should probably be included in future simulations of high-performance running. 

In the final experiment, we tested whether the state prior is useful in motion reconstruction tasks. Thus, we reconstructed three motions from \citet{nitschkeChangeDirection3D2023a} from three marker sets: the full set of 42 markers, a leg set of 11 markers on the pelvis, knees, and ankles, and a distal set of 8 markers on the ankles and hands. We found that adding the prior improves reconstruction accuracy for the two sparse marker sets, but not for the full marker set. This is consistent with our expectations, that the prior is most useful when the tracking term underdetermines the motion, and that there is little benefit or even misregularization when the tracking term already constrains the motion well. Surprisingly, the prior also improves reconstruction of a cutting manoeuvre and curved running, which are motions that the prior has not seen during training, indicating that it can generalise past the specific tasks it was trained on. However, sparse reconstructions still lack accuracy, as all sparse tracking problems introduce a backward trunk lean at the end of the motion that is only partially corrected by the prior. 

Effort or metabolic cost objectives could also be considered priors in the sense that they encode a belief about what governs the recorded movement~\cite{ackermannOptimalityPrinciplesModelbased2010a,crowninshieldPhysiologicallyBasedCriterion1981,rasmussenMuscleRecruitmentMin2001}. 
If the prior does not encode a valid belief about the task, it will bias the solution toward an unwanted result. For example, our state prior is trained on mostly gait data and would bias towards gait-like motions when reconstructing out-of-distribution motions such as dancing or swimming. A similar misalignment occurs between effort minimization  and marker tracking in the sparse reconstruction task. 
Effort minimization biases towards less mechanical work, i.e., shortening the path of the centre of mass, whereas marker tracking aims to reproduce the recorded motion. 
However, cutting maneuvers are not energy-optimal. When the model assumes they are, it produces unwanted artifacts like the backward lean at the end of the movement.
On the other hand, the first two experiments constrain the solution space to periodic gait, which reduces the solution space to a region where the used metabolic cost objective cannot lead to low-energy shortcuts, and thus the misalignment between the objective and prediction task is not an issue.

\begin{figure}[t!]
  \centering
  \includegraphics{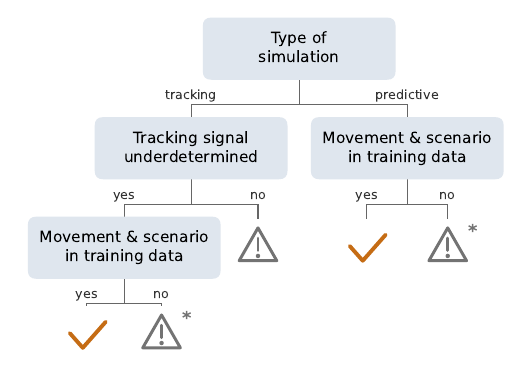}
  \caption{\textbf{Flowchart: When to use or not use a state prior in optimal control simulations.} We generally recommend using a state prior when the task is underdetermined, such as in predictive simulations or sparse reconstruction and if the motion of interest is similar to the training data of the prior. Asterisks indicate that the prior can be used, but it needs to be carefully evaluated for the specific task.}
  \label{fig:prior-flowchart}
\end{figure}

Consequently, we introduce recommendations for when to use a state prior in optimal control simulations (Fig.~\ref{fig:prior-flowchart}). When performing predictive simulations of healthy gait, the prior should be used because it effectively guides the solution towards human-like gait. When performing predictive simulations of non-gait motions, the prior should not be used, as it will bias the solution toward gait-like motions. However, if there is limited data available for the motion of interest, it might be beneficial to fine-tune the prior. For non-healthy gait, the prior could be used with caution, as a bias toward healthy gait is expected, and fine-tuning on impaired gait data is recommended. For this scenario, we have shown that we can reconstruct key features of a gait impairment (Trendelenburg gait), but the choice of objective function might need to be adjusted to account for other factors than metabolic cost, such as impairment-related stability or pain.
For reconstruction tasks, the prior should be used when the tracking term underdetermines the motion, such as with IMU data, while caution applies for non-gait motions. If the motion is well-constrained by the tracking term, such as with a full-body marker set and force plates, the prior only encodes available data and does not improve reconstruction accuracy.

The state prior takes joint angles, velocities, and external forces as input, but not joint torques or muscle activations. We chose this state representation as it consists of quantities that can easily be measured or estimated from experimental data, and that are independent of the musculoskeletal model. Joint torques and muscle activations, on the other hand, depend on the musculoskeletal model and its parameters, which are uncertain and specific to one model. Including them in the state representation would make the prior less generalizable across different musculoskeletal models. We tested including joint torques in the state representation when simulating with generic musculoskeletal models, and found that it slightly improved kinematic accuracy. Leaving out velocities or external forces, however, noticeably reduced accuracy. Thus, we chose joint angles, velocities, and external forces as the most balanced state representation.

We evaluate the prior as the negative log-likelihood of the reconstructed state, which is a measure of how likely the prior considers the reconstructed state to be. The main benefit of this approach is that the prior yields a unitless cost per channel, which is important as the state representation contains different units whose orders of magnitude differ. In comparison, a fixed-weight mean-squared error would require tuning a relative weight for each by hand. Additionally, the heteroscedastic (per-channel) variance learned by the prior accounts for different noise levels in the experimental input data. Thus, the choice of negative log-likelihood as the evaluation metric for the prior's state representation is a practical one. 

While we consider the prior to be a useful tool for predictive simulations and motion reconstruction, our simulation results are not without limitations. A special emphasis needs to be placed on musculoskeletal modelling, for example by enhancing the realism of the ground contact model. Very importantly, a major problem in musculoskeletal modelling is the bias toward male morphology, which is present in the inertial and muscle parameters of the musculoskeletal model. Female musculoskeletal models are emerging lately in the literature and should be considered in any future work~\cite{umeharaSexDifferencesThreeDimensional2026,stansfieldFemalePopulationAveragedMusculoskeletal2026,vanderkrukBIASMECHANICSDoesUnconscious2025a,gambietzBodyHullsMusculoskeletal2026}. Further future work should also investigate how impaired gait and assistive devices can be incorporated into prior-based simulations. 

In conclusion, we have shown that a learned state prior should be used in predictive simulations of human gait, as it guides the solution towards human-like gait and thus improves realism. The state prior can directly be integrated into healthy gait simulations, for example in performance optimization studies, which is the field where we see the most immediate application of our work. However, with cautious fine-tuning of the prior using limited data, this framework may also be extended to impaired gait simulations, and thus enables new applications in rehabilitation and assistive device design. 

%TC:ignore
\section{Methods}
% I've actually written this myself xD

\subsection*{Learned State Prior}

We model the distribution of states real humans occupy during gait with a variational autoencoder (VAE), trained once and then held fixed as a regularisation objective inside the trajectory-optimisation solver (Fig.~\ref{fig:vae-method}).

\textbf{State representation.} Each training example is a single time instant described by $x = (q, \dot q, F)$: the generalised joint coordinates, their velocities, and the external (ground-reaction) forces acting on the body. We chose this representation because $q$ and $F$ are quantities the motion-capture and force-plate data constrain or measure directly, whereas muscle activations and joint torques are not measured and depend on musculoskeletal-model parameters that are themselves uncertain; including them would let modelling error leak into the prior. The prior is therefore agnostic to the muscle model and activation strategy, and can be combined with any of them.
% DICUSSION note: different formulations were tested, and including moments

\textbf{Training data.} The VAE is trained on the AddBiomechanics dataset 1.0 \citep{werlingAddBiomechanicsDatasetCapturing2025}, a large, publicly aggregated motion-capture corpus, from which we used all trials that have synchronised force-plate data. This amounts to ca. \num{58} hours of motion-capture data.
 We represent the skeleton with 23 of the 33 degrees of freedom (DOF) of the \emph{runMaD} musculoskeletal model \cite{nitschkeEfficientTrajectoryOptimization2020d}, excluding the metatarsal and subtalar joints, which were locked in the dataset, and the six pelvis DOF. The pelvis DOF are not included because movements are invariant to global position and heading. None of the reference data shown throughout the results section are part of the AddBiomechanics dataset.

\textbf{Architecture.} The VAE consists of an encoder and a decoder with two hidden layers each. Rather than reconstructing the input directly, the decoder outputs a per-dimension mean and variance, i.e.\ a Gaussian distribution over the reconstructed state as this makes the reconstruction loss unitless, and leads to measurement channels where the experimental data is noisy (e.g.\ joint velocities) being penalised less than channels where the data is precise (e.g.\ joint coordinates). Trajectory optimisation requires continuous gradients, so we use smooth GELU activations and RMSNorm layer normalisation throughout. 

\textbf{Training objective.} The decoder's Gaussian output is trained with the corresponding Gaussian negative log-likelihood (GNLL) against the encoder's input $x$. The training objective is the sum of the GNLL and the Kullback-Leibler (KL) divergence, $\mathcal L = \frac{1}{\beta}\mathrm{GNLL} + \,\mathrm{KL}\!\left(\mathrm{Enc}_\phi(x)\,\|\,\mathcal N(0, I)\right)$, with the KL term up-weighted to $\beta=10$ relative to the reconstruction term, because the GNLL term takes systematically larger values than a conventional mean-squared-error loss would.

The architecture is similar to VPoser \citep{SMPL-X:2019}, a learned pose prior from computer vision, but differs in state content by including velocities and ground-reaction forces, in its probabilistic decoder, and in the smoothness of its activation functions  for use inside an optimal-control solver. Full architecture and training hyperparameters (layer widths, latent dimensionality, optimiser, learning rate, batch size, training length, and data split) are listed in Supplementary Tables~S1 \& S2.

\subsection*{Integration into predictive simulations}

\textbf{Evaluating the prior.} During trajectory optimisation, we optimise a trajectory of states $x(t)$ that must satisfy the musculoskeletal dynamics, and we evaluate the prior at each collocation node. Therefore, the prior is evaluated as the reconstruction term of the VAE using the current state $x$ as input to the encoder and decoder rather than a decoder-only evaluation of a latent variable $z$ as it would be in a typical computer-graphics application \cite{SMPL-X:2019,rempeHuMoR3DHuman2021a}. The prior objective is then the average GNLL across all collocation nodes:
\begin{equation}
  \mathcal J_\mathrm{prior} = \frac{1}{N}\sum_{i=1}^{N} \mathrm{GNLL}\!\left(\mathrm{Dec}_\theta(\mathrm{Enc}_\phi(x_i))\right),
\end{equation}

\textbf{Optimal control problem.} We formulate the trajectory-optimisation problem with direct collocation and implicit dynamics, following \cite{nitschkeEfficientTrajectoryOptimization2020d,koelewijnBioMACSimToolboxMATLABToolbox2025b}, and solve it with IPOPT \cite{wachterImplementationInteriorpointFilter2006a}. 
To avoid biasing the solution toward a specific gait, we initialise the optimisation with a simulation of a standing pose that itself was initialised with a randomised joint configuration. 
Depending on the experiment, we either prescribe a target speed, a target duration, or leave both unconstrained. The optimisation objective is a weighted sum of the prior and a task-specific objective, such as metabolic cost (Experiment 1), effort (Experiment 2), or data tracking and effort (Experiment 3). In addition, arm torques are minimised according to \citep{nitschkeEfficientTrajectoryOptimization2020d}, and we used a soft constraint on the number of ground contact events to avoid solutions where the simulation would yield multiple gait cycles within the loose duration bounds that we set. An overview of the objective terms and their weights is given in Supplementary Table~S3.

\textbf{Musculoskeletal model.} Simulations use the \emph{runMaD-SIPP} model, a three-dimensional full body musculoskeletal model with 92 lower-limb muscles and torque-actuated arms. \emph{runMaD-SIPP} is an adaptation of the 33-DOF \emph{runMaD} model implemented in the BioMAC-Sim-Toolbox \citep{koelewijnBioMACSimToolboxMATLABToolbox2025b}. It combines a symmetrised parameterisation of the generic body-segment inertial parameters from SIPP \cite{gambietzBodyHullsMusculoskeletal2026} with the ground-contact model of \citet{falisseModelingToesContributes2022} that represents foot-ground contact as an exponential (Hunt-Crossley-type) spring-damper rather than the linear spring-damper used in the original \emph{runMaD} model.

\subsection*{Experiments for predictive gait simulations}

\textbf{Experiment 1: Predictive simulations across walking and running speeds.} Here, we want to demonstrate that the prior improves predictive simulations across the full speed range. We impose speeds of 0.73 to 3.53 m/s in 0.1 m/s increments, and 3.53 to 5.53 m/s in 0.2 m/s increments, and compare the resulting kinematics and kinetics to experimental data from Fukuchi et al.\ \citep{fukuchiPublicDatasetRunning2017} (running) and Le Goff et al.\ \citep{legoffPublicDatasetTreadmill2026} (walking), as well as known relationships between speed, cadence, and duty factor \citep{hansenRoleStrideFrequency2017,leacoxEffectRunningSpeed2025}. 
We benchmark our results to the previous state-of-the-art baselines, listed in Table~\ref{tab:baselines}. 
Eligible methods must generate gait without measured gait data as input. They may be trained on experimental data, but none receive a reference motion at prediction time. This excludes imitation-based approaches that track a prescribed motion \cite{liEmbodiedAIMuscleMimic2026,nitschkeEfficientTrajectoryOptimization2020d,leeScalableMuscleactuatedHuman2019,parkBidirectionalGaitNetBidirectional2023}. Among the eligible methods, we benchmark against the best-performing one per methodological category, selected by the agreement of the reported joint kinematics with experimental data. Further studies in each category exist (Optimal \& Reflex Control: \cite{falisseRapidPredictiveSimulations2019d,songNeuralCircuitryThat2015}, Data-driven: \cite{rempeHuMoR3DHuman2021a}, Reinforcement Learning: \cite{suSimulatingHumanWalking2023,schumacherEmergenceNaturalRobust2025a,kimFreeMuscoMotionFreeLearning2025,parkMuscleSynergyPriors2026,weiScalingWholeBodyHuman2026,pengAMPAdversarialMotion2021,badieBioinspiredMorphologyTask2025}). %, but report kinematics that deviate more strongly from experimental reference data or their output cannot be constrained to a specific speed.
As the metric, we use the Wasserstein distance to compare the distributions of joint angles and ground reaction forces between the simulations and the experimental data. We chose this metric as experimental gait data is a spectrum and we also punish simulations that produce only a single gait pattern. 
Thus, this metric is disadvantageous for our method, as the simulations come from a single generic musculoskeletal model and their spread reflects different local minima due to different initial guesses, not the inter-subject variation of the reference data.

\begin{table}[h!]
  \centering
  \caption{State-of-the-Art baselines for predictive simulations. Our comparable baselines are PredSim (PS~\citep{dhondtPredSimFrameworkRapid2024a}), Generative GaitNet (GGN~\citep{parkGenerativeGaitNet2022}), GaitDynamics (GD~\citep{tanGaitDynamicsGenerativeFoundation2026}), and GaitEncoder (GE~\citep{magruderGaitEncoderFoundationModel2026}). Methods are ranging from purely physics-based to purely data-driven, with our method being a hybrid of both.}
  \label{tab:baselines}
  \begin{tabular}{l l l l}
    \toprule
    Baseline & Method & Type & Ref.\\
    \midrule
    Ours & Optimal control & Physics+Data & -\\
    PS & Optimal control & Physics-based & \citep{dhondtPredSimFrameworkRapid2024a}\\
    GGN & RL & Physics+Data & \citep{parkGenerativeGaitNet2022}\\
    GD & Diffusion & Data-driven & \citep{tanGaitDynamicsGenerativeFoundation2026}\\
    GE & Autoencoder & Data-driven & \citep{magruderGaitEncoderFoundationModel2026}\\
    \bottomrule
  \end{tabular}
\end{table}

% weiScalingWholeBodyHuman2026 might be better than GGN. 
We also run walking simulations at unconstrained speeds, where neither speed nor cadence is prescribed, which would not yield realistic gait with previous optimal control formulations. In these simulations, we use the metabolic costs calculated with Bhargava's model \ \citep{bhargavaPhenomenologicalModelEstimating2004a} as the task-specific objective. The Bhargava model uses muscle excitations as inputs, which we replaced with muscle activations because optimising excitations leads to non-physiological solutions \cite{vandenbogertStrangeEffectsActivation2025}. Metabolic costs are calculated as the energy expenditure rate divided by the body mass and gait speed, yielding a cost of transport in J/(kg m). In free-speed conditions, the usual practice of optimising (squared) metabolic cost directly would lead to bad conditioning as the inverse of speed creates large gradients for low speeds that quickly vanish for higher speeds. We therefore use the logarithm of the cost of transport as the objective, which yields a more uniform gradient across speeds. All simulations for Experiments 1 \& 2 were discretised with 50 collocation nodes per half-gait cycle, and we reran each simulation scenario until 10 converged solutions were found, to account for the stochasticity of the initial guess.

\textbf{Experiment 2: Hypothesis testing.} We demonstrate that the prior enables hypothesis testing in gait. The first hypothesis we test is which objective function best explains human gait: reducing energy costs or avoiding fatigue. The idea of energy cost minimisation, e.g.\ a metabolic cost objective, stems from the observation that humans tend to select a walking pattern that minimises energy expenditure \cite{zarrughOptimizationEnergyExpenditure1974b}. Fatigue avoidance has been suggested as an alternative hypothesis \cite{ackermannOptimalityPrinciplesModelbased2010a}, which is correlated to the sum of cubic muscle activations \cite{crowninshieldPhysiologicallyBasedCriterion1981a}. Here, we compare the differences in metabolic cost profiles and ground reaction forces between simulations using either of the two objective functions. The second hypothesis we test is which metabolic cost model best explains human gait. We compare four widely used models: Bhargava et al.\ \cite{bhargavaPhenomenologicalModelEstimating2004a}, Umberger et al.\ \cite{umbergerStanceSwingPhase2010b}, Houdijk et al.\ \cite{houdijkEvaluationHillBased2006a}, and Lichtwark et al.\ \cite{lichtwarkModifiedHillMuscle2005a} and run simulations at unconstrained speeds to see how the choice of metabolic model affects gait. 
Thirdly, we test the effect of weak hip abductors on gait. We run simulations at a speed of 1.33 m/s with the metabolic cost objective and reduce the maximum isometric force of the gluteus medius, gluteus minimus, and tensor fasciae latae muscles to 100\%, 50\%, 30\%, 20\%, 15\%, and 10\% of their nominal strength. 
Lastly, we test the effect of ground contact stiffness and damping on running gait. We run simulations at a speed of 4.5 m/s with three different ground contact models: the default Hunt-Crossley model, a four-times-stiffer model, and a model without damping, i.e. ideal energy return, while using effort as the task-specific objective. We compare the resulting kinematics, ground reaction forces, and metabolic costs to see how the ground contact model affects running gait.

\textbf{Experiment 3: Sparse tracking.} Current biomechanical reconstruction methods rely on a large number of markers to reconstruct full-body kinematics, which increases experiment set-up times. Here, we show that the prior enables sparse tracking and reconstruction of full-body kinematics from a small number of markers. We re-run the example simulations from Nitschke et al.\ \cite{nitschkeChangeDirection3D2023a}, where they reconstruct straight running, curved running, and a cutting maneuver from a full set of markers. We repeat the same simulations with two reduced marker sets: (i) only markers on the legs, and (ii) only markers on the ankles and hands. We compare the resulting kinematics to the full-marker reconstruction and sparse reconstruction without the prior. 

%TC:endignore

%TC:ignore
\section*{Data \& Code Availability}
Code: \url{https://github.com/gambimar/BiomechPriorVAE}. Simulation results underlying the figures will be deposited publicly before submission.

\section*{Acknowledgments}
This work was funded by the Deutsche Forschungsgemeinschaft (DFG, German Research Foundation) -- Project-ID 442419336 -- CRC 1483 (EmpkinS).

\section*{Conflict of Interest}
The authors declare no competing interests.

\section*{Author Contributions}
% use the CRediT taxonomy (https://casrai.org/credit/) to describe each author's contributions
\textbf{Conceptualization:} MG, ZZ;
\textbf{Methodology:} MG, ZZ, TB, AK;
\textbf{Software:} MG, ZZ;
\textbf{Validation:} MG, TB, XW;
\textbf{Formal analysis:} MG;
\textbf{Investigation:} MG, ZZ, TB, XW;
\textbf{Data curation:} MG, ZZ, XW;
\textbf{Writing - original draft:} MG;
\textbf{Writing - review \& editing:} MG, TB, AK;
\textbf{Visualization:} MG;
\textbf{Supervision:} AK.

\bibliographystyle{unsrtnat-etal}
\bibliography{biomechpriorVAE}
%TC:endignore
\end{document}